arXiv 2026

# On four representations of the law of aftershock evolution

A.V. Guglielmi, O.D. Zotov

*Schmidt Institute of Physics of the Earth RAS, Moscow, Russia*

*guglielmi@mail.ru, ozotov@inbox.ru*

## Abstract

The relaxation of the earthquake source following the main shock has been investigated both theoretically and experimentally, based on data regarding the evolution of aftershocks. It is shown that Omori’s picture of aftershock evolution is not holistic, whereas the Hirano-Utsu representation contradicts observations. Linear and discrete representations are proposed. They are based on original concepts regarding deactivation and the proper time of the source. The difference between the linear and discrete representations lies in the procedure for processing experimental data on aftershocks. In the linear representation, proper time is calculated based on the frequency of aftershocks. In a discrete representation, the source's proper time is measured by an imaginary underground clock, the rate of which differs radically from that of universal time. In the linear representation, the frequency of aftershocks is described by the simplest nonlinear differential equation. In the discrete representation, the aftershock frequency is described by a first-order linear differential equation. It has been found that aftershock activity decays exponentially with the source's proper time.

## Introduction

In the late 19th century, Omori mathematically formulated the law governing the decay of aftershock activity, reflecting the relaxation process of the source following the main earthquake shock [Omori, 1894]. He introduced the concept of the aftershock frequency $n(t)$ as a smooth function of universal time, and this long determined the main direction of theoretical research on aftershocks. An alternative approach involves timing aftershocks not according to universal time, but according to the earthquake source's proper time. The flow of the source's proper time may, generally speaking, differ from the flow of universal time. It turned out that the use of proper time makes it possible to develop an original methodology for identifying previously unknown properties of aftershock evolution.

We will analyze four representations of the law governing aftershock evolution: the hyperbolic representation of Omori [Omori, 1894]; the power-law representation of Hirano and Utsu [Hirano, 1924; Jeffreys, 1938; Utsu, 1861, 1962; Utsu, Ogata, Matsu'ura, 1995]; the linear representation of Guglielmi, Zavyalov, and Zotov [Guglielmi, 2016, 2017; Faraoni, 2020; Zavyalov, Guglielmi, Zotov, 2020; Guglielmi, Zotov, 2021; Zavyalov et al., 2022; Guglielmi et al., 2023, 2025a,b]; and the discrete representation of Zotov and Guglielmi [Zotov, Guglielmi, 2025a,b; Guglielmi, Zotov, 2025]. We will find that the hyperbolic representation is non-holistic, while the power-law representation is irrelevant. Linear and discrete representations are holistic and relevant. They are based on the concepts of deactivation and the proper time of the earthquake source. The difference between the linear and discrete representations lies in the methods used to calculate the source's proper time. A distinctive feature of the discrete representation is that proper time is measured in terms of underground tremors, which are viewed as the ticking of a peculiar kind of "underground clock". Both methods of aftershock timing yield interesting new results in a most unexpected way.

## Hyperbolic representation

Omori's law entered modern geophysical literature in the author's original formulation:

$$n(t) = \frac{k}{c+t}. \qquad (1)$$

Here, $k > 0$, $c > 0$, $t \geq 0$ (e.g., see [Guglielmi, 2017]). The parameter $k$ characterizes the earthquake source. The choice of parameter $c$ is determined by the experimenter. For example, one can adopt the rule of choosing $c$ such that the frequency $n(0)$ coincides with the peak aftershock frequency, or of selecting both parameters $k$ and $c$ so that the smooth function (1) provides the best approximation of the discrete aftershock sequence. Thus, Omori's law is a one-parameter law: the source is characterized by the parameter $k$ = const.

According to Omori's law, the decay of aftershock activity continues indefinitely. Does the relaxation of the source transition into a regime of steady-state background activity? We will not dwell on this interesting and complex question. The complexity arises from the fact that the decaying sequence of aftershocks is often interrupted by a new mainshock, followed by a series of new aftershocks.

## Power representation

Thirty years after the formulation of Omori's law, Hirano noted that function

$$n(t) = \frac{k}{(c+t)^p} \qquad (2)$$

provides a better fit to the observational data than function (1) [Hirano, 1924]. Thus, Hirano introduced a second phenomenological parameter of the source: the exponent $p$

The significance of power-law representation in the approximation of experimental data was not immediately appreciated. It was only through the work of Utsu [Utsu, 1961, 1962; Utsu, Ogata, Matsu'ura, 1995] that formula (2) came to be widely used by aftershock researchers. It is often referred to as the Omori-Utsu law, although it would be more accurate to speak of the Hirano-Utsu law. The compact and reasonably accurate representation of aftershock sequences using the parameters $p$ and $k$ has become firmly established in seismology (see, e.g., [Ogata, Zhuang, 2006; Rodrigo, 2015; Salinas-Martínez et al., 2023; Bogomolov, Rodkin, Sychev, 2025]).

Experience has shown that the parameter $p$ varies from case to case, roughly within the range of 0.7–1.5, but averages 1.1 that is it differs only slightly from the integer value $p = 1$. This prompts us to look at the history of physics. Let us recall that both Cavendish and Coulomb, in their investigations of forces, obtained a considerable scatter of experimental data points; yet, without hesitation, they favored a law in which the force decreases with distance from a spherical source according to an integer power. Should we not also disregard the scatter of the data points and set p = 1 (thereby returning to the elegant hyperbolic Omori law) while simultaneously raising the question of further experimental validation of the law through more refined observations of aftershocks? In the next section, we describe the methodology for processing and analyzing aftershocks and present a test result that proved to be highly surprising.

In concluding this section, we would like to dispel the naive impression regarding the stimulus that prompted us, ten years ago, to analyze and critique the hyperbolic and power-law representations of aftershocks. The impetus was the consideration mentioned above, and by no means a desire to revise established views at the expense of the truth. We were guided by the principles of transcendental phenomenology, bearing in mind that Coulomb's law and the law of Universal gravitation were grounded in experience (specifically as phenomenological laws) long before the formulation of electrodynamics and the general theory of relativity, within which both laws appear as logical consequences of the fundamental equations of macroscopic physics. A fundamental theory of earthquakes has not yet been formulated, so we have no way to derive the laws of aftershock evolution deductively. However, we can draw upon historical experience and the general principles of transcendental phenomenology.

## Linear representation

Formally, Omori's law (1) can be viewed as the graph of a smooth function representing a harmonic progression. The formula for the terms of the harmonic progression

$$n_i = \frac{1}{A + (i-1)B},$$

where $i$ = 1, 2, 3, …, suggests the idea of a simple substitution of the dependent variable: $n \to 1/n$. As a result, without loss of generality, we obtain a linear law of aftershock evolution. It represents a graph of a discrete sequence of terms of an arithmetic progression.

We will show that, in the linear representation, the evolution law admits a far-reaching generalization. Indeed, let us introduce a deactivation coefficient $\sigma(t)$ that provides a generalized description of the state of the source as a dynamic system. We postulate the law of aftershock evolution in the form of the simplest differential equation

$$dg/dt = \sigma(t), \quad (3)$$

in which

$$g(t) = \frac{n(0) - n(t)}{n(0)n(t)}.$$

It is easy to verify that for $\sigma = \mathrm{const}$ the solution to equation (3) is equivalent to Omori's law (1). However, the dynamic equation (3) is remarkable precisely because, unlike the classical Omori law, it imposes no restrictions on the time dependence of the deactivation coefficient. At $\sigma = \sigma(t)$ Omori's hyperbolic representation is severely violated.

The hyperbolic representation can be reconstructed if the proper time of source

$$\tau(t) = \int_0^t \sigma(t')dt'$$

is used instead of universal time. The dynamics of aftershocks is now described by the elegant equation

$$\frac{dn}{d\tau} + n^2 = 0, \quad (4)$$

which follows directly from our postulate (3). The solution has the form

$$n(\tau) = \frac{n_0}{1 + n_0 \tau}. \quad (5)$$

Thus, we retain Omori's hypothesis regarding the hyperbolic decay of aftershock activity, but only under the indispensable condition that we use the source's proper time rather than universal time.

We believe that in time it will be possible to calculate the deactivation coefficient deductively. However, at the current stage of the theory's development, we will approach the problem in reverse and learn to measure the deactivation coefficient experimentally. In doing so, we will take a necessary and significant step toward improving methods for the experimental study of earthquake sources.

This was the underlying concept of the phenomenological theory of aftershocks presented in the works [Guglielmi, 2016, 2017; Zavyalov, Guglielmi, Zotov, 2020; Zavyalov et al., 2022; Guglielmi et al., 2023]. The objective was to determine the law governing aftershock evolution without imposing constraints on the time dependence of the deactivation coefficient. The law in the mathematically equivalent forms (3)–(5) was established under the additional condition that, given a constant deactivation coefficient, the general law reduces to the classical Omori law (1). The additional condition is imposed based on the consideration that, on average across numerous measurements, the exponent $p$ is close to unity.

The question naturally arises for the experimenter: How can the ideas of deactivation and proper time be practically applied in processing and analyzing an aftershock sequence? To do this, it is necessary to synchronize universal time with one's proper time, that is, to find the function $\tau(t)$. Thus, the solution reduces to experimentally determining the deactivation coefficient $\sigma(t)$ by solving the inverse source problem.

Let us rewrite equation (4) in the form

$$\frac{dn}{dt} + \sigma n^2 = 0 , \qquad (6)$$

and formulate the inverse source problem, the essence of which is as follows. We need to determine the deactivation coefficient based on the aftershock frequency known from observations. Omitting the technical details, which we have described repeatedly in the aforementioned literature, we present the correct solution to the inverse problem:

$$\sigma(t) = \frac{d}{dt}\langle g(t)\rangle. \qquad (7)$$

The angle brackets here denote a regularization procedure that amounts to smoothing the auxiliary function $g(t) \propto 1/n(t)$, which we obtained by processing the aftershock sequence.

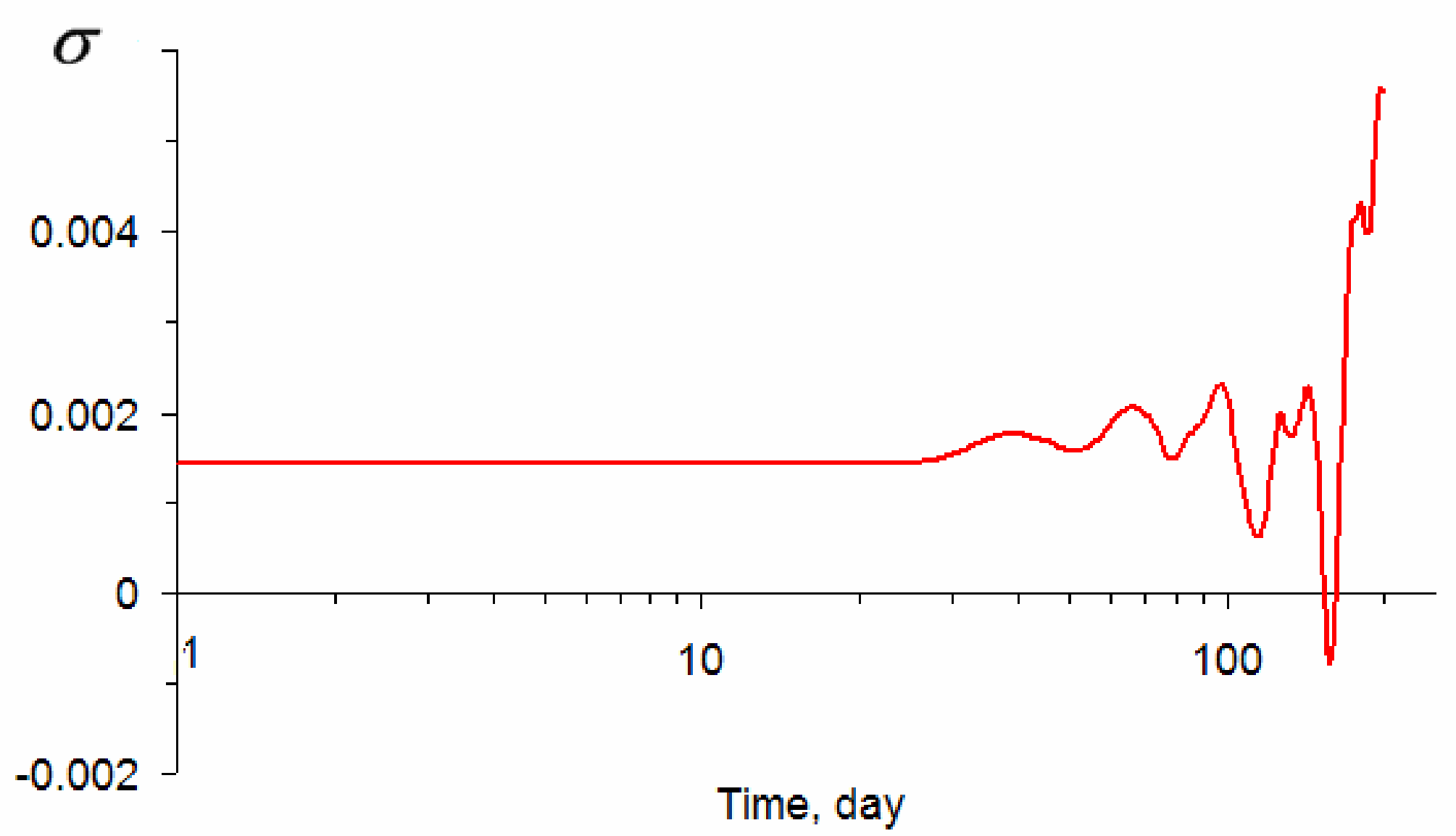

**Fig. 1**. Variation of the deactivation coefficient of the Tohoku earthquake source.

Figure 1 shows the solution to the inverse problem for the Tohoku earthquake [Zotov, Guglielmi, 2025]. The main shock, with a magnitude of M = 9.1, occurred on March 11, 2011, off the east coast of Honshu Island, Japan. We draw particular attention to the fact that the deactivation coefficient is much less than unity. A simple explanation for this can be given (see Appendix).

We see from the figure that $\sigma \neq \text{const}$ at the beginning of the evolution. This means that Omori's law holds. We named the first stage of evolution the Omori epoch. In this specific case, the duration of the Omori epoch is approximately one month. The deactivation coefficient is $\sigma = 0.0014$. In the Omori epoch, the evolution of aftershocks proceeds in a quite predictable manner.

The first stage culminates in a bifurcation, followed by the second stage of evolution. It is characterized by the chaotic, unpredictable behavior of the deactivation coefficient. Thus, Omori's law is valid, but it is not holistic.

We processed more than a hundred events in a similar manner. In all cases without exception, we observed a two-stage relaxation regime for the source. In the first stage, $\sigma$ = const. Upon the conclusion of the Omori epoch, a bifurcation occurs, and in the second stage of evolution, the deactivation coefficient changes unpredictably over the course of universal time. It should be emphasized that law (4) holds over a wide range of mainshock magnitudes; moreover, the deactivation coefficient during the Omori epoch decreases monotonically as magnitude increases.

The simple yet radical idea of shifting from the selection of empirical formulas describing aftershock activity to an experimental study of the source as a dynamic system whose state is characterized in a generalized way by a deactivation coefficient proved to be productive. However, it must be emphasized that the success of our methodology was largely ensured by the prior work of seismologists on modeling aftershock sequences using the Hirano-Utsu fitting formula. On average, based on numerous measurements, the parameter $p$ in formula (2) differs only slightly from unity. Given the exceptionally large sample size, it can be stated with confidence that the mean value of $p = 1.1$ was measured with high precision. A slight deviation of $p$ from unity can be attributed to an unaccounted-for systematic measurement error. This observation determined our choice of postulate (3), which is equivalent to the dynamic equations (4) and (6).

We have established that the relaxation of the source proceeds in two stages, with Omori's law being satisfied during the first stage. The question arises as to whether the Hirano-Utsu law holds during the second stage of relaxation? We arrive at a negative answer through the following reasoning.

Calculating the deactivation coefficient using formula (2) leads us to the conclusion that the quantity $\sigma(t)$ must be a monotonic function of time. Meanwhile, experience shows that the deactivation coefficient depends non-monotonically on universal time during the second stage of the source's evolution.

## On foreshocks

Aftershocks are part of the so-called classical triad, which consists of a sequence of foreshocks, the main shock, and aftershocks [Guglielmi, 2015; Guglielmi, Zotov, 2024]. It turns out that the auxiliary function $g(t) = 1/n(t)$, whose derivative equals the deactivation coefficient, is useful to employ in the processing and analysis of foreshocks.

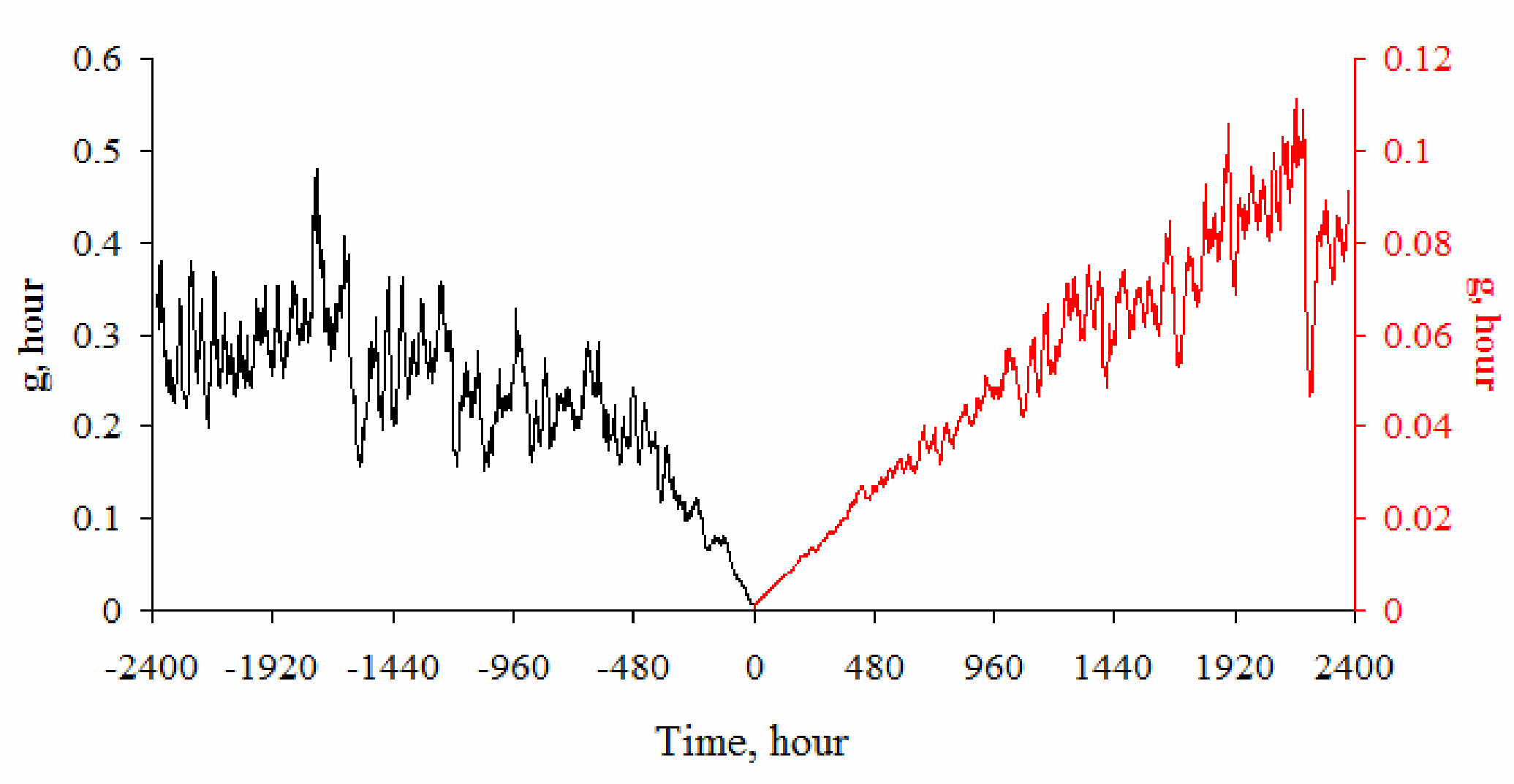

**Fig. 2**. Averaged dynamics of foreshocks and aftershocks. Time zero corresponds to the moment of the main shock with magnitude $M \geq 5$.

Indeed, let us turn to Figure 2. It shows the function $g(t)$ for foreshocks and aftershocks. The graph was constructed using the superposed epoch method based on data from the USGS/NEIC catalog (https://www.usgs.gov). Over the period from 1973 to 2019, 4,205 complete classical triads were identified, comprising both foreshocks (20,783) and aftershocks (123,566). For each main shock, a circular epicentral zone was determined using the methodology described in the paper [Zavyalov and Zotov, 2021]. Earthquakes with focal depths of up to 250 km were considered. The time interval for analysis spanned 2,400 hours before and 2,400 hours after the main shock. Events were accumulated in hourly bins, followed by 25-point smoothing and calculation of the function $g(t)$. The deactivation coefficient $\sigma^+$ can be calculated from the right branch of the function $g(t)$ shown in Figure 2. Similarly, the value of $\sigma^-$ can be calculated using the left branch. It is natural to call $\sigma^-$ the source activation coefficient.

Let us note the characteristic bend in the function $g(t)$ in the interval from 800 to 1200 hours prior to the main shock. The inflection reflects the source's departure from the stationary background activity regime and the onset of intensification preceding the main shock. In connection with the mention of background seismic activity, it is worth noting that in the steady-state regime, the proper time of the source effectively comes to a halt, $\tau = 0$.

## Discrete representation

The transition from the hyperbolic Omori representation to the linear representation does not overcome the difficulty inherent in the concept of the instantaneous aftershock rate $n(t)$. Even allowing for the well-known caveat that we measure frequency over time intervals that are physically, and by no means mathematically infinitesimally small, a sense of dissatisfaction remains because we fail to take into account the evident discreteness of underground tremors. To remedy the aforementioned shortcoming, we explored various approaches and conceived the idea of using the seismic shocks themselves to mark the discrete proper time of the source. The transition from a continuous to a discrete representation of aftershock laws deprives us of powerful mathematical analysis methods, however, at the experimental data preprocessing stage, our approach to this transition proved exceptionally productive.

So, the idea is to use underground tremors as the ticking of an imaginary subterranean clock, counting off the source's proper time. We want to synchronize the underground clocks with the clocks showing universal time. Let us introduce a coordinate plane ($x$, $t$), where $x$ is a non-negative real number and $t$ is universal time. At the points $x_i = i$, $t_i(x_i)$, where $i = 1, 2, 3, \ldots$,, we plot the aftershock excitation times $t_i$ listed in the representative earthquake catalog. A multitude of points on the coordinate plane makes it possible to synchronize the discrete proper time of the source with world time.

We expect that, in the Omori epoch, the time intervals $T_i = t_{i+1} - t_i$ between successive aftershocks form a geometric progression. Let us briefly outline the guiding considerations in this regard. It follows from the evolution equation (4) that the interval $T = 1/n$ is a linear function of the continuous proper time of the source, $\tau$. The function $T(\tau)$ represents the graph of an arithmetic progression. To fully grasp the essence of the matter, we need only recall the relationship between arithmetic and geometric progressions. Without dwelling further on elementary arguments, we state immediately that the smooth synchronization function $t(x)$ representing the graph of the corresponding geometric progression must be an exponential function. In the discrete representation, the deactivation coefficient is calculated using formula

$$\sigma(x) = \frac{d}{dx} \ln t(x). \qquad (8)$$

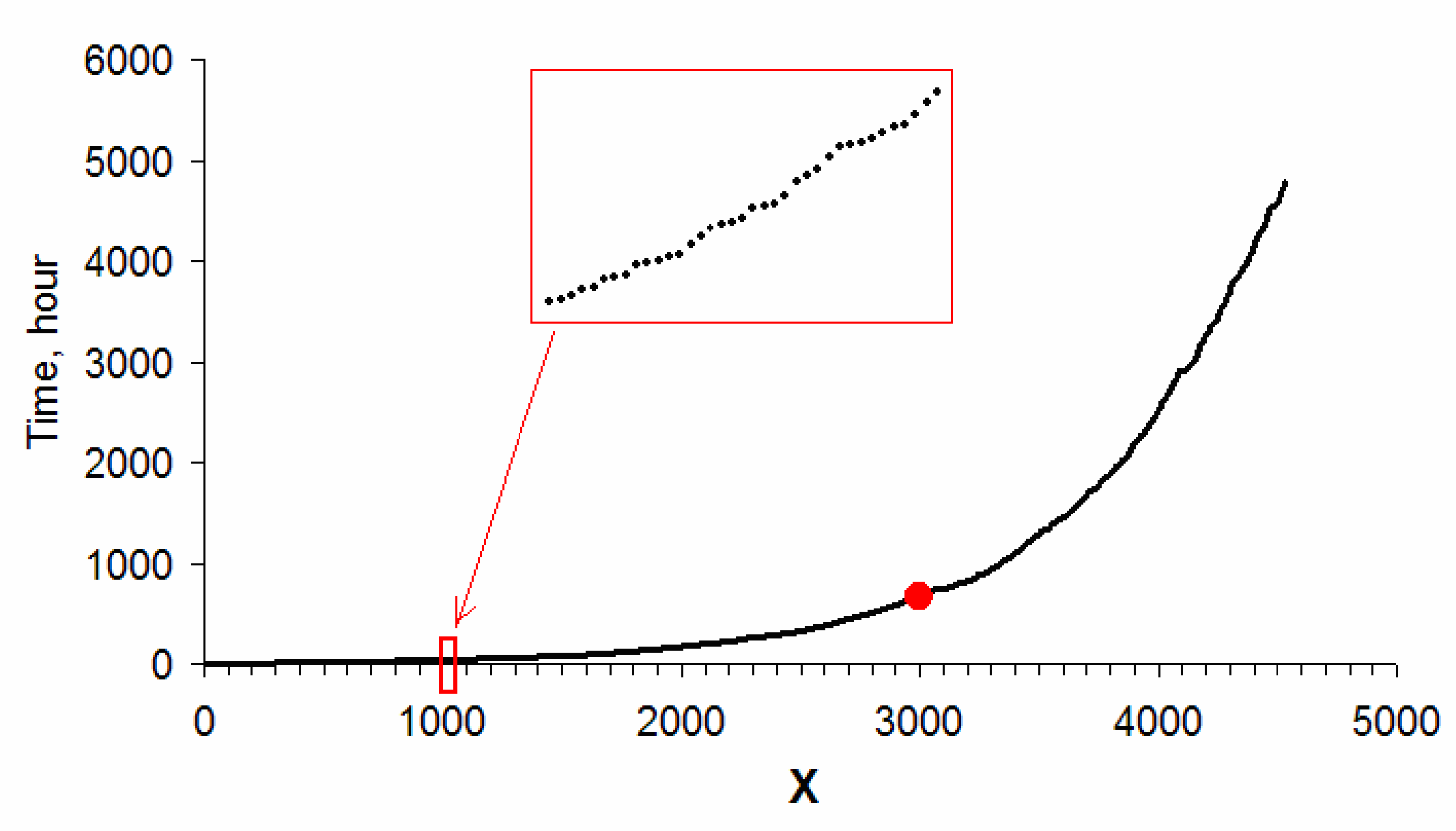

**Fig. 3**. Synchronization of the Tohoku earthquake source's proper time (horizontal axis) and Universal Time (vertical axis). The points on the coordinate plane ($x$, $t$) represent the excitation times of 4,537 aftershocks. The red dot marks the end of the Omori epoch. A special window shows a fragment of the set of points on an enlarged scale.

We present an observation that brilliantly confirmed our expectations: an analysis of the aftershocks of the Tohoku earthquake [Zotov, Guglielmi, 2025]. We number the aftershock sequence using the natural numbers $i = 1, 2, 3, \ldots$, determine $t_i$ from the USGS/NEIC catalog, and plot the corresponding points on the coordinate plane. The result is shown in Figure 3. The points merged into a continuous line. The set of points is well approximated by the exponential function

$$t(x) = t_1 \exp\left[\sigma(x-1)\right], \quad x \geq 1, \qquad (9)$$

with $t_1 = 7.2$ h and $\sigma = 0.0014$. The coefficient of determination is close to unity: $R^2 = 0.97$. The proper functioning of our special-purpose clock is evidenced by the fact that the deactivation coefficient $\sigma$ = 0.0014 matches the value we obtained from the linear representation of the Tohoku aftershock evolution.

We see that the concept of proper time enriches the arsenal of methods for the experimental study of aftershocks. Having determined the synchronization function from observations, we can calculate the focus deactivation coefficient using formula (7) where $g(t) = T_0 + \tau(t)$ if a linear representation is used, or using the formula

$$\sigma(x) = \frac{d}{dx}\langle \ln t(x) \rangle,$$

if a discrete representation is used.

To conclude this section, let us state in words the law of aftershock evolution in a discrete representation: *The interval between successive aftershocks increases exponentially over proper time.* Accordingly, the frequency of aftershocks decreases exponentially with the passage of proper time:

$$n(\chi) = n_0 \exp(-\chi), \quad \chi(x) = \int_1^x \sigma dx. \qquad (10)$$

This means that the evolution of aftershocks in a discrete representation is described by a linear differential equation.

$$\frac{dn}{d\chi} + n = 0. \qquad (11)$$

In light of the foregoing, the outlines of a future theory of aftershocks are emerging. We base the theory on three postulates:

1. There exists a function representing the state of the earthquake source as a dynamic system undergoing relaxation following the main shock.

2. There exists a proper time of the source, defined as a functional of the state function.

3. Aftershock activity decays exponentially with the source's proper time.

The system of axioms remains incomplete, since we have not yet found a dynamic equation for the state function.

## Discussion

Hyperbolic and power-law representations are widely used in the experimental study of aftershocks. Conversely, linear and discrete representations are not used and are virtually unmentioned in the literature. The reason for this likely lies in the fact that the significance of the deactivation coefficient and proper time for earthquake physics is not yet fully recognized and understood. The high effectiveness of the concepts of deactivation and proper time in the search for new properties and evolutionary patterns of the source has not been fully appreciated. By tradition, the Omori $k$ parameter and the Hirano-Utsu $p$ parameter continue to be widely used in the processing and analysis of aftershocks.

We must recognize that the theory of aftershocks, based on linear and discrete representations, is phenomenological in nature. The deactivation coefficient and the time synchronization function are introduced as postulates based on heuristic considerations. We are unable to derive the postulates from the laws of general physics. It is solely the success in identifying new properties of aftershocks that motivates us to employ linear and discrete representations of source relaxation following the main shock.

The discovery of two substantially different stages of source relaxation is undoubtedly significant. In the first stage, Omori's law is strictly obeyed, as evidenced by the constancy of the deactivation coefficient. Upon completion of the first stage, a phenomenon resembling a bifurcation occurs in the source region, and the deactivation coefficient begins to vary non-monotonically with universal time $t$. It is remarkable that the hyperbolic Omori law continues to hold, provided only that we replace absolute time $t$ with the source's proper time $\tau$. The exponential growth of the synchronization function $t(x)$ when using underground clocks appears exceptionally interesting. We introduced the term "underground clock" as an evocative figure of speech, but a look at Figure 3 reveals that a strictly ordered process unfolds within the colossal rock mass following the formation of a major discontinuity.

Centuries of searching for methods to measure time led to the invention of remarkable instruments – sundials, water clocks, and mechanical, quartz, and electronic watches. Culminating in modern times with the creation of atomic clocks that measure time with unprecedented precision. In light of this, our proposal to measure the timing of dynamic processes within an earthquake source using a "subterranean clock" might at first glance appear to be an unacceptable anachronism, but only at first. It should be noted that underground clocks belong to the class of special-purpose clocks. Furthermore, experience has shown that underground clocks can be synchronized with clocks indicating Universal Time. Finally, timing based on underground clocks revealed properties and patterns of earthquakes that had been overlooked when using clocks displaying Universal Time. Let us illustrate what has been said with a concrete example.

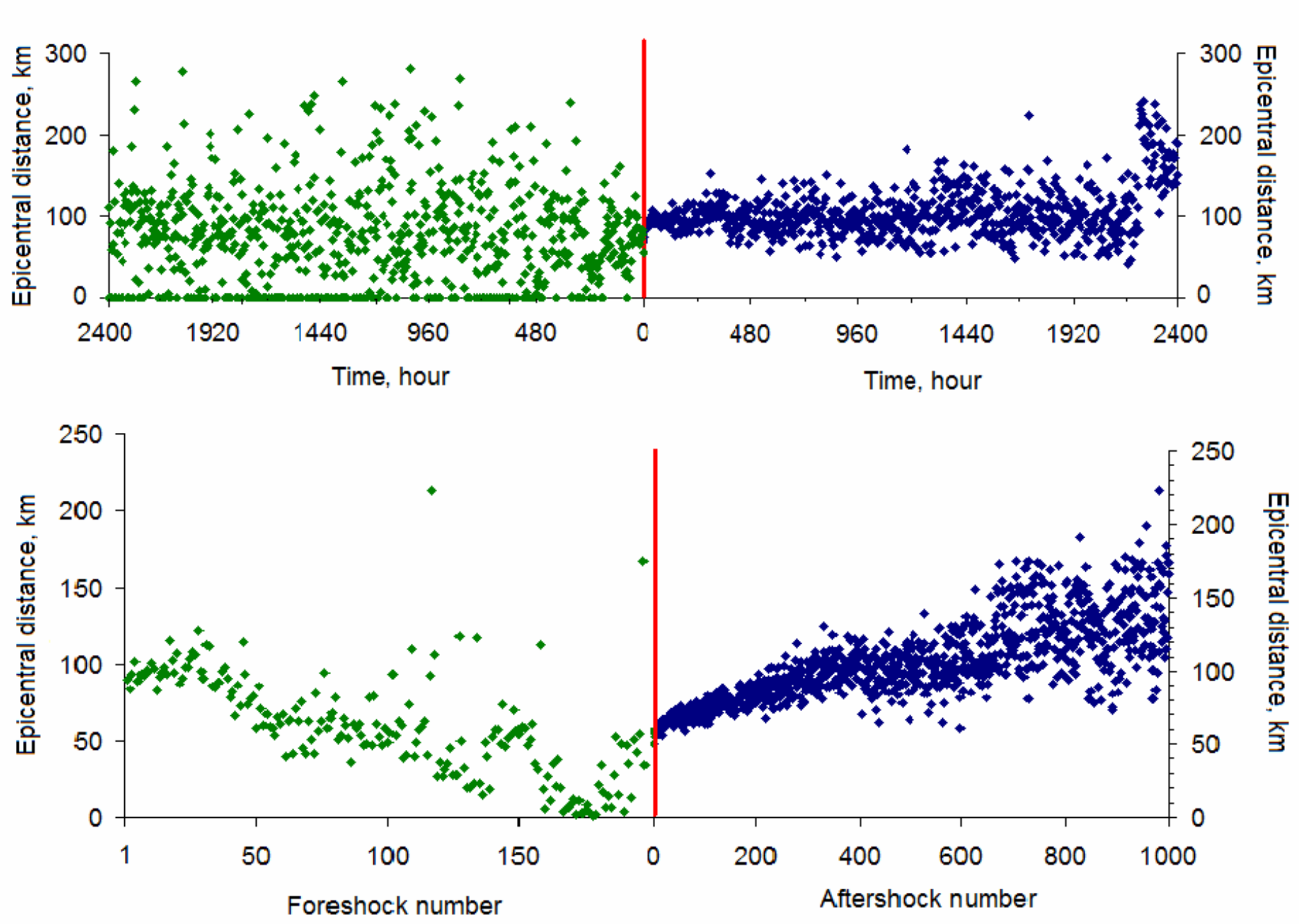

**Fig. 4**. Convergence of foreshocks and divergence of aftershocks. Results of a statistical study on the dependence of the epicentral distance of foreshocks (left) and aftershocks (right) on universal time (top panel) and on the discrete proper time of the source. Point 0 corresponds to the time of the main shock.

In Figure 4, the distances of foreshocks and aftershocks from the mainshock epicenter are plotted along the vertical axes. Universal Time is plotted along the horizontal axes on the upper panel, and source proper time on the lower panel. The figure is based on data from the USGS/NEIC catalog. Between 1973 and 2019, we identified 138 main shocks with magnitudes of $M \geq 7.5$. All foreshocks and aftershocks within a $\pm 100$ days interval relative to the main shock in the main shock's epicentral zone were taken into account. The hypocentral depths of the foreshocks, the main shock, and the aftershocks do not exceed 250 km. A total of 1,960 foreshocks and 34,627 aftershocks were observed. The striking contrast between the upper and lower panels in Figure 4 convincingly demonstrates the effectiveness of our proper-time chronometry method in studying the dynamics of tectonic earthquake sources.

## Conclusion

We have outlined two doctrinally distinct approaches to the study of aftershocks following a strong earthquake. The traditional approach consists in selecting empirical formulas that describe the evolution of aftershocks. The one-parameter Omori law and the two-parameter Hirano-Utsu law are derived using this approach. The alternative approach is based on a new idea. Alternative approach aims to investigate the earthquake source as a dynamic system whose state is described by a single parameter - the deactivation coefficient. Aftershock data are used to calculate the deactivation coefficient.

An alternative approach led to the formulation and theoretical incorporation of an original concept regarding the proper time of an earthquake source. The flow of proper time may differ from the flow of universal time. Using the source's proper time, we formulated a holistic law of aftershock evolution. A bifurcation of the source has been identified in the linear representation of the evolution, while an exponential decay of aftershock activity over the course of the source's proper time has been observed in the discrete representation.

***Acknowledgments***. We would like to express our gratitude to A.D. Zavyalov for many years of collaboration on the study of aftershocks and for the valuable assistance provided regarding the issues addressed in this paper. This work was supported by the Ministry of Science and Higher Education of the Russian Federation within the framework of state assignments of the Schmidt Institute of Physics of the Earth of the Russian Academy of Sciences.

**Appendix**

The first postulate of the classical theory of aftershocks states:

*The aftershocks frequency n(t) is a smooth slowly decaying function of time.*

We shall show that the smallness of the deactivation coefficient compared to unity follows directly from this. Indeed, the slow variation of the frequency implies that the quantity $n$ changes only slightly over the interval $T = 1/n$. Mathematically, this is expressed by the strong inequality

$$n \gg \left| \frac{d}{dt} \ln n \right| . \qquad \text{(A.1)}$$

Let $s$ be a small dimensionless parameter, $0 < s << 1$. We can now rewrite (A.1) in the form

$$\frac{dn}{dt} + sn^2 = 0 . \qquad \text{(A.2)}$$

Equation (A.2) coincides with the dynamic equation (6) if we identify the small parameter with the deactivation coefficient. Thus, the holistic law of aftershock evolution and the smallness of the deactivation coefficient logically follow from the postulate stated above. The experiment showed that $\sigma \ll 1$ in all the cases examined. This supports the choice of the fundamental postulate of aftershock theory in its linear representation.